\documentclass[prd,twocolumn,showpacs,preprintnumbers,amsmath,amssymb,superscriptaddress,floatfix,nofootinbib]{revtex4-2}
\usepackage{float}
\usepackage{graphicx}
\usepackage{bm}
\usepackage{amsmath}
\usepackage{amsfonts}
\usepackage{amssymb}
\usepackage{color}
\usepackage{multirow}
\usepackage{soul}

\usepackage[colorlinks, citecolor=blue,anchorcolor=red,menucolor=red, linkcolor=red,filecolor=red,runcolor=red,urlcolor=blue,frenchlinks=red]{hyperref}

\begin{document}
\title{Mass spectra and decay properties of the higher excited $\rho$ mesons}

\author{Xue-Chao Feng}
\affiliation{College of Physics and Electronic Engineering, Zhengzhou 
University of Light Industry, Zhengzhou 450002, China}

\author{Zheng-Ya Li}
\affiliation{School of Physics and Microelectronics, Zhengzhou University, Zhengzhou, Henan 450001, China}

\author{De-Min Li}\email{lidm@zzu.edu.cn}
\affiliation{School of Physics and Microelectronics, Zhengzhou University, Zhengzhou, Henan 450001, China}

\author{Qin-Tao Song}\email{songqintao@zzu.edu.cn}
\affiliation{School of Physics and Microelectronics, Zhengzhou University, Zhengzhou, Henan 450001, China}

\author{En Wang}\email{wangen@zzu.edu.cn}
\affiliation{School of Physics and Microelectronics, Zhengzhou University, Zhengzhou, Henan 450001, China}

\author{Wen-Cheng Yan}\email{yanwc@zzu.edu.cn}
\affiliation{School of Physics and Microelectronics, Zhengzhou University, Zhengzhou, Henan 450001, China}

\begin{abstract}
Although there are some experimental hints for the higher  excited $\rho$ mesons, our  knowledge of their properties is much poor theoretically. Based on our
recent work about excited $\rho$ mesons [Phys.Rev.D104(2021)034013], we present the mass spectra and decay properties of the higher excited $\rho$ mesons with the 
modified Godfrey-Isgur quark model and the  $^3P_0$ strong decay model, and compare our predictions with the experimental hints, which should be helpful to search for these higher excited $\rho$ mesons.
\end{abstract}

\date{\today}

\maketitle

\section{Introduction}
\label{sec1}

In the last decades, a large number of light vector mesons with mass below 2.4~GeV have been reported experimentally~\cite{BESIII:2018ldc,BaBar:2007ceh,BESIII:2020xmw,BESIII:2019apb,BaBar:2019kds,BESIII:2018zbm,Zyla:2020zbs}, and a lot of research works have been 
conducted theoretically based on these experimental data~\cite{Masjuan:2013xta,Feng:2021igh,Wang:2020kte,Pang:2019ttv,Pang:2019ovr,Zhao:2019syt}. These experimental and theoretical  studies are crucial to deepen our understanding of the light vector mesons. In addition, the higher excited light vector mesons above 2.4~GeV play an important role in the processes involving the baryon anti-baryon interactions~\cite{Xiao:2019qhl}.
However, there are few theoretical and experimental studies on the higher excited light vector mesons, so it is necessary to study their properties, which should be helpful to search for them experimentally.

Recently, we  have studied the mass spectra and 
strong decay properties of the excited $\rho$ mesons $Y(2040)/\rho(2000)$, $\rho(1900)$, and $\rho(2150)$ with the modified Godfrey-Isgur quark model (MGI) including the screening  
effects
and the $^3P_0$ strong decay model, and found that they can be assigned as 
 $\rho(2^3D_1)$, $\rho(3^3S_1)$, and $\rho(4^3S_1)$, respectively~\cite{Li:2021qgz}. We have also shown the screening effects are crucial to promote theoretical descriptions of the masses of the observed excited $\rho$ mesons, as shown in Fig.~2 of Ref.~\cite{Li:2021qgz}.
 We predict that the $Y(2040)$ mainly decays to  $\pi\pi$ and $\pi\omega$, which is  in good agreement with the measurements of the processes $p\bar{p}\to \pi\pi$ and $e^+e^-\to \omega\pi$~\cite{Hasan:1994he,BESIII:2020xmw}. For $\rho(1900)$, the predicted main decay modes are $\pi\omega$ and $a_2(1320)\pi$, which is also in consistent with the experimental measurements~\cite{Zyla:2020zbs}. 
For $\rho(2150)$, its main decay modes are expected to be $\pi\pi$, $\pi\omega(1420)(\to 6\pi)$, and $\pi a_2(1320)(\to 6\pi)$, consistent with the experimental results~\cite{Zyla:2020zbs}.  Based on the successful description of the excited mesons $Y(2040)/\rho(2000)$, $\rho(1900)$, and $\rho(2150)$, we would like to extend  
our study  to the higher excited $\rho$
 mesons with masses above 2.4~GeV in this work.

Although the higher excited $\rho$ mesons above 2.4~GeV  have not been reported experimentally, there exist some hints of them in the experimental measurements. For instance, in 2002, Anisovich {\it et. al} analyzed experimental data of the process  $p\bar{p}\rightarrow\omega\eta\pi^0$ measured by the Crystal Barrel Collaboration, and showed that there exists a peak structure with $J^{PC}=1^{--}$ around 2.4~GeV~\cite{Anisovich:2002su}. Later in 2006, the {\it BABAR} Collaboration  reported the $\omega\pi^+\pi^-\pi^0$ energy-dependent reaction cross section, as shown in Fig.~18 of Ref.~\cite{BaBar:2006vzy}, where the blue data points correspond to the cross section  data  of the process  $e^+e^-\rightarrow\omega\pi^+\pi^-\pi^0$.
From Fig.~18 of Ref.~\cite{BaBar:2006vzy}, one can easily find  a peak around 2.6~GeV, which may be the higher excited $\rho$ meson according to the $G$ parity conservation.


In 2007, the {\it BABAR} Collaboration  presented the cross section of the process $e^+e^-\rightarrow K^+K^-\pi^+\pi^-\pi^0$, 
where there is  a peak structure around 2.5 GeV~\cite{BaBar:2007qju}. The analysis of the {\it BABAR} measurements  within the Vector Meson Dominance model shows the existences of the $\rho(1900)$ and another resonance with mass of $M=2550\pm13$~MeV and width of $\Gamma=209\pm26$~MeV~\cite{Lichard:2018enc}, where the
latter one is expected to have the same quantum numbers as $\rho(1900)$. In addition, the cross  sections of $e^+e^-\rightarrow\eta'\pi^+\pi^-$ and $e^+e^-\rightarrow f_1(1285)\pi^+\pi^-$ processes reported by {\it BABAR} 
indicate that there exist
 some structures around 2.5~GeV, and it is difficult to claim 
the existences of new   resonances
 due to the insufficient data statistics~\cite{BaBar:2007qju}. 

In a word, there exist some hints of the higher excited $\rho$ mesons
according to the above experimental measurements, as pointed out in Ref.~\cite{Wang:2021abg}, and the existences of the higher excited $\rho$ mesons
above 2.4~GeV need to be confirmed by more precise 
experimental measurements in future. The theoretical predictions for their mass spectra and decay properties are 
crucial  to search for them experimentally. Thus, we will study the mass spectra and decay properties of the higher excited $\rho$ mesons above 2.4~GeV with the MGI model and $^3P_0$ strong decay model, as used in our previous work~\cite{Li:2021qgz}

%

The organization of this paper is as follows. First we introduce the MGI model and the $^3P_0$ model in Sec.~\ref{sec2} and Sec.~\ref{sec3}, respectively. Then we present the numerical results of the mass spectra and strong decays for the higher excited $\rho$ mesons in Sec.~\ref{sec4}. Finally a short summary is given in Sec.~\ref{sec5}.

\section{Relativistic Quark Model}
\label{sec2}
The Godfrey-Isgure relativistic quark model (GI model) was proposed in 1985 by Godfrey and Isgur~\cite{Godfrey:1985xj}, and is widely used in describing the mass spectra of mesons~\cite{Pang:2017dlw,Lu:2014zua,Lu:2016bbk}, especially for the 
lower excited mesons. However, the masses of the higher  excited mesons measured by experiments are much lower than the predictions of the GI model. Many studies show that the screening effects play an important role in studying the higher radial and orbital excited mesons~\cite{Chao:1992et,Ding:1993uy}. First we present a brief introduction of the GI model, and the MGI model that includes the screening effects.

\subsection{GI model}

Within the GI model, the Hamiltonian of the meson's internal interaction can be written as~\cite{Godfrey:1985xj},
\begin{equation}\label{hamtn}
  \tilde{H}=\sqrt{m_1^2+\mathbf{p}^2}+\sqrt{m_2^2+\mathbf{p}^2}+\tilde{V}_{\mathrm{eff}}(\mathbf{p,r}),
\end{equation}
where $m_1$ and $m_2$ denote the masses of quark and antiquark in the meson, 
respectively, $\mathbf{p}=\mathbf{p}_1=-\mathbf{p}_2$ is the  center-of-mass momentum, and $\mathbf{r}$ corresponds to the spatial coordinate.  The $\tilde{V}_{\mathrm{eff}}(\mathbf{p,r})$ is the effective potential between quark and antiquark, and can be expressed as follows in the non-relativistic limit,
\begin{equation}
V_{\mathrm{eff}}(r)=H^{\mathrm{conf}}+H^{\mathrm{hyp}}+H^{\mathrm{so}}\label{1},
\end{equation}
where $H^{\mathrm{conf}}$ is spin-independent potential, includes the spin-independent linear confinement and Coulomb-type interaction,
\begin{align}
H^{\mathrm{conf}}&=\Big[-\frac{3}{4}(c+br)+\frac{\alpha_s(r)}{r}\Big] \bm{F}_1\cdot\bm{F}_2  \nonumber\\
  &=S(r)+G(r)\label{2},
\end{align}
where $\langle \bm{F}_1\cdot\bm{F}_2 \rangle =-4/3$ for a meson. $\alpha_s(Q^2)$ is the running coupling constant of QCD, which depends on the energy scale $Q$. $\alpha_s(Q^2)$ is divergent at low $Q$ region. The authors of Ref.~\cite{Godfrey:1985xj} assume that $\alpha_s(Q^2)$ saturates as $\alpha_s(Q^2=0)=\alpha_s^{\mathrm{critital}}$, which is regarded as a parameter to be determined by fitting to experimental data. $\alpha_s(r)$ is obtained from $\alpha_s(Q^2)$ by using Fourier transform, where $r$ is the relative distance between quark and antiquark.
$H^{\mathrm{hyp}}$ denotes the color-hyperfine interaction, 
\begin{align}
H^{\mathrm{hyp}}&=-\frac{\alpha_s(r)}{m_{1}m_{2}}\Bigg[\frac{1}{r^3}\Big(\frac{3\bm{S}_1\cdot\bm r \bm{S}_2\cdot\bm r}{r^2} -\bm{S}_1\cdot\bm{S}_2\Big)\nonumber\\
  &+\frac{8\pi}{3}\bm{S}_1\cdot\bm{S}_2\delta^3 (\bm r)\Bigg]  \bm{F}_1\cdot\bm{F}_2 .
\end{align}
$H^{\mathrm{so}}$ is the spin-orbit interaction which contains the color-magnetic term $H^{\mathrm{so(cm)}}$ and the Thomas-precession term $H^{\mathrm{so(tp)}}$,
\begin{align}
H^{\mathrm{so}}=H^{\mathrm{so(cm)}}+H^{\mathrm{so(tp)}}, \label{3.2}
\end{align}
\begin{align}
H^{\mathrm{so(cm)}}=\frac{-\alpha_s(r)}{r^3}(\frac{1}{m_2}+\frac{1}{m_2})(\frac{\bm{S}_1}{m_1}+\frac{\bm{S}_2}{m_2})\cdot\bm{L}(\bm{F}_1\cdot\bm{F}_2)
\end{align}
\begin{align}
H^{\mathrm{so(tp)}}=-\frac{1}{2r}\frac{\partial H^\mathrm{conf}}{\partial r}(\frac{\bm{S}_1}{m_1^2}+\frac{\bm{S}_2}{m_2^2})\cdot \bm{L}.
\end{align}
The spin-orbit interaction will give rise to the  mixing between spin singlet $ ^1L_J$ and spin triplet $ ^3L_J$  if $m_1 \neq m_2$.

In GI model, the relativistic effects are introduced by two main ways.
Firstly, in the quark-antiquark scattering, the interactions should depend on both quark momentum $\vec{p}_1$ and antiquark momentum $\vec{p}_2$, or a linear combination of them as $\vec{p}_1-\vec{p}_2$ and $\vec{p}_1+\vec{p}_2$, so they must be nonlocal interaction potentials as pointed out in Ref.~\cite{Godfrey:1985xj}. In order to take this effect into account,
a smearing {function} $\rho_{12} \left(\mathbf{r}-\mathbf{r}^{\prime}\right)$ is used to transform the basic potentials $G(r)$ and $S(r)$ into the smeared ones $\tilde{G}(r)$ and  $\tilde{S}(r)$, the specific form is
\begin{eqnarray}\label{sme}
\tilde{f}(r)=\int d^3r'\rho_{12}(\mathbf{r}-\mathbf{r'})f(r'),
\end{eqnarray}
and the smearing function $\rho_{12}(\mathbf{r}-\mathbf{r'})$ is defined as
\begin{align}
\rho_{12}\left(\mathbf{r}-\mathbf{r'}\right)=\frac{\sigma_{12}^3}{\pi ^{3/2}}e^{-\sigma_{12}^2\left(\mathbf{r}-\mathbf{r'}\right)^2},
\end{align}
\begin{align}
\sigma_{12}^2=\sigma_0^2\Bigg[\frac{1}{2}+\frac{1}{2}\left(\frac{4m_1m_2}{(m_1+m_2)^2}\right)^4\Bigg]+s^2\left(\frac{2m_1m_2}{m_1+m_2}\right)^2.\label{si12}
\end{align}

 For a heavy-heavy $Q\bar{Q}$ meson system, $\rho_{12}(\mathbf{r}-\mathbf{r'})$ will turn into delta function $\delta^3(\mathbf{r}-\mathbf{r'})$ as one increases the quark mass $m_Q$, in this case, one can obtain $\tilde{f}(r)=f(r)$, which indicates that the relativistic effects can be neglected for a heavy-heavy $Q\bar{Q}$ meson. However, the relativistic effects are important for heavy-light mesons and light mesons, therefore, it is necessary to consider relativistic effects in the study of excited $\rho$ mesons.

Secondly, the momentum-dependent factors are introduced to modify the effective potentials, as shown in the following formula,
\begin{eqnarray}
 \tilde{G}(r)&\to \left(1+\frac{p^2}{E_1E_2}\right)^{1/2}\tilde{G}(r)\left(1 +\frac{p^2}{E_1E_2}\right)^{1/2},\\ \nonumber\\
 \frac{\tilde{V}_i(r)}{m_1m_2}&\to \left(\frac{m_1m_2}{E_1E_2}\right)^{1/2+\epsilon_i}\frac{\tilde{V}_i(r)}{m_1 m_2}\left(\frac{m_1 m_2}{E_1 E_2}\right)^{1/2+\epsilon_i},
\end{eqnarray}
where $\tilde{G}(r)$ is the Coulomb-type potential, and $\tilde{V}_i(r)$ {represents} the contact, tensor, vector spin-orbit and scalar spin-orbit terms as explained in Ref.~\cite{Godfrey:1985xj}. In the non-relativistic limit, those momentum-dependent factors will become unity.

{With the help of the detailed expressions of 
Eq.\,(\ref{hami})},
 we obtain the mass spectrum and wave functions of mesons
 by  solving the Schr\"odinger equation with the Gaussian expansion method, the meson wave functions are used as inputs to investigate the subsequent strong decays for mesons.

\subsection{MGI model involving the screening effects}
 
 In GI model, linear potential $br$ tends to infinity with the increasing of the distance between quark and antiquark, however for the large distance,  the quark-antiquark pairs are generated in the vacuum, 
 and the colors of quark and antiquark are screened  in a meson, which implies that the simple linear potential $br$ cannot accurately describe the interaction between quarks for 
higher excited mesons~\cite{Chao:1992et,Ding:1993uy,Song:2015nia,Feng:2022esz}. The study of the lattice QCD  shows that the effective potential of 
mesons
  deviates from the traditional linear potential when the distance between quark and antiquark is greater than 1~fm~\cite{Born:1989iv}. Therefore, it is very important to introduce the screening  effects when studying the masses of higher radial and orbital excited mesons.

The screening effects are introduced to 
 the GI model by substituting linear potential $br$ 
with the following form~\cite{Chao:1992et,Ding:1993uy,Song:2015nia},
\begin{eqnarray}
V(r)=br\rightarrow V^{\mathrm{scr}}(r)=\frac{b(1-e^{-\mu r})}{\mu},
\end{eqnarray}
 when $r$ is small enough, we can have $V^{\mathrm{scr}}(r)=V(r)$, therefore, this replacement will not affect the 
lower-lying 
  meson states.
The parameter $\mu$ is related to the strength of the screening effects and can be determined by fitting to the experimental data.

Furthermore, the smeared potential of Eq.(\ref{sme}) can be rewritten as~\cite{Song:2015fha,Song:2015nia},
\begin{eqnarray}
\tilde V^{\mathrm{scr}}(r)&=&\frac{b}{\mu r}\Bigg[e^{\frac{\mu^2}{4 \sigma^2}+\mu r}\Bigg(\frac{1}{\sqrt{\pi}}\int_0^{\frac{\mu+2r\sigma^2}{2\sigma}}e^{-x^2}dx-\frac{1}{2}\Bigg)\nonumber\\
&&\times\frac{\mu+2r\sigma^2}{2\sigma^2}+r-e^{\frac{\mu^2}{4\sigma^2}-\mu r}\frac{\mu-2r\sigma^2}{2\sigma^2}\nonumber\\
&&\times\Bigg(\frac{1}{\sqrt{\pi}}\int_0^{\frac{\mu-2r\sigma^2}{2\sigma}}e^{-x^2}dx-\frac{1}{2}\Bigg)\Bigg], \label{Eq:pot}
\end{eqnarray}
 where $\sigma=\sigma_{12}$ defined in Eq.~(\ref{si12}).

  The MGI model is widely used in the studies
  of mass spectrum of heavy-heavy mesons, heavy-light meson and light mesons~\cite{Pang:2018gcn,Li:2021qgz,Pang:2017dlw, Pang:2019ttv,Song:2015nia,Song:2015fha,Wang:2019mhs,Wang:2018rjg,Hao:2019fjg,Li:2022ybj,Gao:2022bsb}. The parameters of the MGI model  are taken from Ref.~\cite{Pang:2017dlw}, 
    as tabulated in Table~\ref{tab:parameter}. 
  The values of other parameters can be found in Ref.~\cite{Godfrey:1985xj}.

   \begin{table}[htpb]
\begin{center}
\caption{ \label{tab:parameter}Parameter values in the MGI model \cite{Pang:2017dlw}.  }
\begin{tabular}{cccc}
\hline\hline
 Parameter                         & Value  &  Parameter  &Value\\
\hline
$m_u$~(GeV) &0.163 &$s$ &1.497 \\
$m_d$~(GeV) &0.163  &$\mu$~(GeV) &0.0635 \\
$m_s$~(GeV) &0.387&$\epsilon_{\rm c}$ &-0.138\\
 $b$~(GeV$^2$) &0.221&$\epsilon_{\rm{sov}}$ &0.157\\
  $c$~(GeV) &-0.240&$\epsilon_{\rm sos}$ &0.9726\\
$\sigma_0$~(GeV) &1.799&$\epsilon_{\rm t}$ &0.893\\
\hline\hline
\end{tabular}
\end{center}
\end{table}
\section{$^3P_0$ strong decay model}

\label{sec3}
 In addition to the mass spectrum, the strong decay properties  are crucial for experiments to search for the higher excited $\rho$ mesons.
Here, we give a brief introduction of the $^3P_0$ model which is widely used in studying two-body OZI-allowed strong decays of mesons~\cite{Xue:2018jvi,Wang:2017pxm,Pan:2016bac,Hao:2019fjg,Feng:2021igh}.

\begin{figure}[htpb]
\includegraphics[scale=0.5]{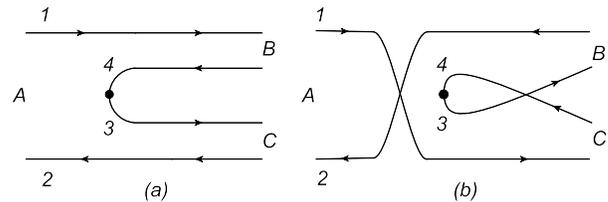}
\vspace{0.0cm}\caption{Two possible diagrams contributing to the process $A\rightarrow B+C$
in the $^3P_0$ model.}\label{fig:feiman}
\end{figure}

 The $^3P_0$ model was originally proposed by Micu~\cite{Micu:1968mk} in 1968, and later it was further developed by Le Yaouanc $et$ $al$~\cite{LeYaouanc:1973ldf,LeYaouanc:1972vsx}. $^3P_0$ model has been considered as an effective tool to study two-body OZI-allowed strong decays of hadrons. For the process $A\rightarrow B+C$, the main idea of $^3P_0$ model is that a flavor-singlet and color-singlet quark-antiquark pair with $J^{PC}=0^{++}$ is created from the vacuum firstly, then, the created antiquark (quark) combines with the quark (antiquark) in meson $A$ to form  meson $B$($C$) as shown in Fig.~\ref{fig:feiman}(a) and Fig.~\ref{fig:feiman}(b).

The transition operator $T$ of the decay $A\rightarrow B+C$ in the $^3P_0$ model is given by \cite{Blundell:1996as},
\begin{eqnarray}
T&=-3\gamma\sum\limits_m \langle 1m1-m|00\rangle\int d^3\boldsymbol{p}_3d^3\boldsymbol{p}_4\delta^3(\boldsymbol{p}_3+\boldsymbol{p}_4)\nonumber\\ \nonumber\\
&\times{\cal{Y}}^m_1\left(\frac{\boldsymbol{p}_3-\boldsymbol{p}_4}{2}\right
)\chi^{34}_{1,-m}\phi^{34}_0\omega^{34}_0b^\dagger_3(\boldsymbol{p}_3)d^\dagger_4(\boldsymbol{p}_4),
\end{eqnarray}
where $\boldsymbol{p}_3(\boldsymbol{p}_4$) is the momentum of the created quark (antiquark). The dimensionless parameter $\gamma$ stands for the strength of the quark-antiquark $q_3\bar{q}_4$ pair created from the vacuum, and  is usually determined by fitting to the experimental data.
$\chi^{34}_{1,-m}$, $\phi^{34}_0$ and $\omega^{34}_0$ are spin, flavor, and color wave functions of the created quark-antiquark pair, respectively.

With the  transition operator $T$, the helicity amplitude ${\cal{M}}^{M_{J_A}M_{J_B}M_{J_C}}(\boldsymbol{P})$ can be written as,
\begin{align}
\langle BC|T|A\rangle &=\delta^3(\boldsymbol{P}_A-\boldsymbol{P}_B-\boldsymbol{P}_C){\cal{M}}^{M_{J_A}M_{J_B}M_{J_C}}(\boldsymbol{P}),
\end{align}
where $|A\rangle$, $|B\rangle$ and $|C\rangle$ denote the mock meson states defined in Ref.~\cite{Hayne:1981zy}, and $\boldsymbol{P}$ is the momentum of meson $B$ in the center of mass frame.
Then the helicity amplitude is obtained as,
\begin{align}
&{\cal{M}}^{M_{J_A}M_{J_B}M_{J_C}}(\boldsymbol{P})\nonumber\\=&\gamma \sqrt{8E_AE_BE_C}\sum_{\renewcommand{\arraystretch}{.5}\begin{array}[t]{l}
\scriptstyle M_{L_A},M_{S_A},\\\scriptstyle M_{L_B},M_{S_B},\\\scriptstyle M_{L_C},M_{S_C},m
\end{array}}\renewcommand{\arraystretch}{1}\!\!
\langle L_AM_{L_A}S_AM_{S_A}|J_AM_{J_A}\rangle\nonumber\\&\times \langle L_BM_{L_B}S_BM_{S_B}|J_BM_{J_B}\rangle \langle L_CM_{L_C}S_CM_{S_C}|J_CM_{J_C}\rangle\nonumber\\&\times\langle 1m1-m|00\rangle \langle\chi^{14}_{S_BM_{S_B}}\chi^{32}_{S_CM_{S_C}}|\chi^{12}_{S_AM_{S_A}}\chi^{34}_{1-m}\rangle\nonumber\\
&\times[f_1I(\bm{P},m_1,m_2,m_3)\nonumber+(-1)^{1+S_A+S_B+S_C}\nonumber\\
&\times f_2 I(-\bm{P},m_2,m_1,m_3)],
\end{align}
where $f_1=\langle\phi^{14}_B\phi^{32}_C|\phi^{12}_A\phi^{34}_0\rangle$ and $f_2=\langle\phi^{32}_B\phi^{14}_C|\phi^{12}_A\phi^{34}_0\rangle$ correspond to the flavor superposition of the two combinations depicted in Fig.~\ref{fig:feiman}(a) and Fig.~\ref{fig:feiman}(b), respectively. The momentum space integral $I(\bm{P},m_1,m_2,m_3)$ is given by
\begin{align}
I(\bm{P},m_1,m_2,m_3)=&\int d^3\bm{p}\psi^*_{n_BL_BM_{L_B}}(\frac{m_3}{m_1+m_3}\bm{P_B}+\bm{p})\nonumber\\ &\times\psi^*_{n_CL_CM_{L_C}}(\frac{m_3}{m_2+m_3}\bm{P_B}+\bm{p})\nonumber\\
&\times\psi_{n_AL_AM_{L_A}}(\bm{P_B}+\bm{p})\mathcal{Y}_1^m(\bm{p}).
\end{align}

Furthermore, the partial wave amplitude ${\cal{M}}^{LS}(\boldsymbol{P})$ for the decay $A\rightarrow B+C$  can be expressed as~\cite{Jacob:1959at},
\begin{align}
{\cal{M}}^{LS}(\boldsymbol{P})=&\sum_{\renewcommand{\arraystretch}{.5}\begin{array}[t]{l}
\scriptstyle M_{J_B},M_{J_C},\\\scriptstyle M_S,M_L
\end{array}}\renewcommand{\arraystretch}{1}\!\!
\langle LM_LSM_S|J_AM_{J_A}\rangle \nonumber\\
&\times\langle J_BM_{J_B}J_CM_{J_C}|SM_S\rangle\nonumber\\
&\times\int d\Omega\,\mbox{}Y^\ast_{LM_L}{\cal{M}}^{M_{J_A}M_{J_B}M_{J_C}}
(\boldsymbol{P}).\label{pwave}
\end{align}

Finally, within the relativistic phase space, the total width $\Gamma(A\rightarrow B+C)$ can be
expressed in terms of the partial wave amplitude squared~\cite{Blundell:1996as},
\begin{eqnarray}
\Gamma(A\rightarrow B+C)= \frac{\pi
|\boldsymbol{P}|}{4M^2_A}\sum_{LS}|{\cal{M}}^{LS}(\boldsymbol{P})|^2, \label{width1}
\end{eqnarray}
where $|\boldsymbol{P}|=\frac{\sqrt{[M^2_A-(M_B+M_C)^2][M^2_A-(M_B-M_C)^2]}}{2M_A}$, $M_A$, $M_B$, and $M_C$ are the masses of the mesons $A$, $B$ and $C$, respectively.

\section{Numerical RESULTS}
\label{sec4}

\subsection{Mass spectrum analysis}

First, we have calculated the masses of the higher excited $\rho$ 
mesons
 within the MGI and GI models,  which  are shown in Table~\ref{tab:mass}. There are six higher excited $\rho$ mesons in the region of  $2.4\sim 3.0$~GeV, which are $\rho(5^3S_1)$, $\rho(6^3S_1)$,  $\rho(7^3S_1)$, $\rho(4^3D_1)$, $\rho(5^3D_1)$, and $\rho(6^3D_1)$. The deviations between the predictions of MGI and GI models are about $300\sim 500$~MeV, which implies the screening potential plays an important role 
in studying
 the masses of the higher excited $\rho$ mesons.

\begin{table}[htpb]
\begin{center}
\caption{\label{tab:mass}Masses of the higher
excited $\rho$ mesons predicted  by the GI and MGI models.}
\begin{tabular}{cccc}
\hline\hline
 State                         & MGI (MeV)            & GI (MeV) &\\
\hline
$\rho(5^3S_1)$ &2542  &2817 & \\
$\rho(6^3S_1)$ &2774  &3160& \\
$\rho(7^3S_1)$ &2967  &3470  & \\
$\rho(4^3D_1)$&2624&2915  &\\
$\rho(5^3D_1)$&2840&3239  &\\
$\rho(6^3D_1)$ &3020 &3554 &\\

\hline\hline
\end{tabular}
\end{center}
\end{table}

\begin{figure}[htpb]
\includegraphics[scale=0.55]{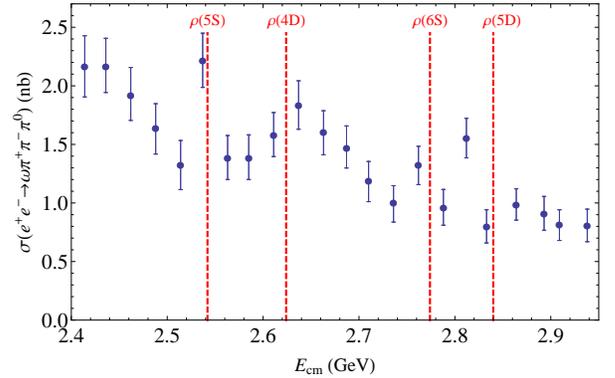}
\vspace{0.0cm}
\caption{Comparison of the masses of the higher excited $\rho$ mesons predicted by MGI model with the cross section of the process $e^+e^-\rightarrow\omega\pi^+\pi^-\pi^0$  reported by {\it BABAR}. The red dotted lines are masses predicted with MGI model, and the blue data points are the {\it BABAR} experimental data.}\label{fig:zhiliang}
\end{figure}

The recent analyses of the {\it BABAR} measurements about the process $e^+e^-\rightarrow K^+K^-\pi^+\pi^-\pi^0$
indicate the existence of one resonance with mass $M=(2550\pm13)$~MeV~\cite{Lichard:2018enc}, which is in good agreement with our predicted mass of $\rho(5^3S_1)$. In addition, the cross section of the process $e^+e^-\rightarrow\omega\pi^+\pi^-\pi^0$ measured by {\it BABAR} indicates that there are  hints of the higher excited $\rho$ mesons. From Fig.~\ref{fig:zhiliang}, one can find one datum point around 2.54~GeV, very close to our predicted mass
 $2542$~MeV of $\rho(5^3S_1)$,  and an enhancement structure around 2.64~GeV, which could be related to state  $\rho(4^3D_1)$ with mass of $2624$~MeV. Another datum point appears around 2.76~GeV, close
to the predicted mass of $\rho(6^3S_1)$. Although some 
hints 
of the higher excited $\rho$ mesons exist, one can not claim those resonances due to
the poor quality of the present data, and the more accurate measurements in future are
crucial to confirm the existence of these resonances.

\begin{figure}[htpb]
\includegraphics[scale=0.55]{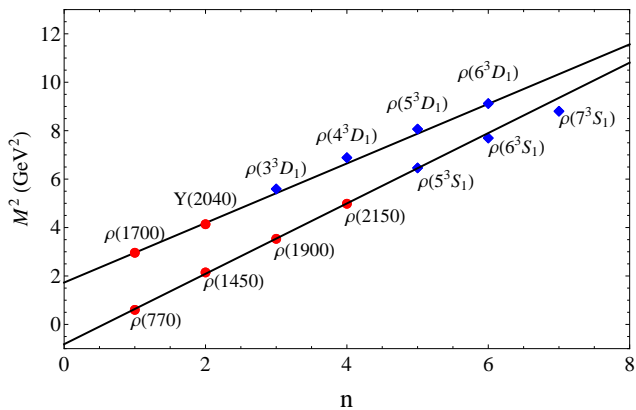}
\vspace{0.0cm}
\caption{The Regge trajectories for the $\rho(n^3S_1)$ and $\rho(n^3D_1)$ masses.}\label{fig:regge}
\end{figure}

In addition, the light mesons with different radial excitation could be well fit to the following quasi-linear $(n, M^2)$ Regge trajectories~\cite{Anisovich:2000kxa},
\begin{equation}
    M^2_n =M^2_0 +(n-1)\mu^2,
\end{equation}
where $M_n$ stands for the mass of the meson with radial quantum number $n$, and $M^2_0$ and $\mu^2$ are the paprameters of the trajectories. With the assignments of $\rho(770)$ as $\rho(1^3S_1)$, $\rho(1450)$ as $\rho(2^3S_1)$,  $\rho(1900)$ as $\rho(3^3S_1)$, $\rho(1700)$ as $\rho(1^1D_1)$, and $Y(2040)$ as  $\rho(2^3D_1)$, respectively, we plot the Regge trajectories in Fig.~\ref{fig:regge}, where one can find that the masses of  $\rho(5^3S_1)$, $\rho(6^3S_1)$,  $\rho(7^3S_1)$, $\rho(4^3D_1)$, $\rho(5^3D_1)$, and $\rho(6^3D_1)$ predicted by MGI model are consistent with the estimations of Regge trajectories.

\subsection{Decay behavior analysis}

 \begin{table}[htpb]
\begin{center}
\caption{ \label{tab:5S}The decay widths of $\rho(5^3S_1)$ (in MeV),  the initial state
mass is set to be 2542~MeV and the masses of the final states are taken from RPP~\cite{Zyla:2020zbs}.}
\begin{tabular}{c|cc|cc}
\hline\hline
 Channel                 & Mode      &$\rho(5^3S_1)$            & Mode           &$\rho(5^3S_1)$   \\
\hline
 $1^-\rightarrow 0^-0^-$      & $\pi\pi$     & 2.52  & $KK$  &$\textless0.01$ \\
                              & $\pi\pi(1300)$  & 12.71 & $KK(1460)$  & 0.01  \\
                               & $\pi\pi(1800)$  &9.28  &  &    \\
  \hline
  $1^-\rightarrow 0^-1^-$     &$\pi\omega$  &0.81        &$KK^*(1680)$ & 0.01 \\
                              &$\rho\eta$  &0.03     &$\rho(1700)\eta$ & 0.09\\
                              &$\omega(1420)\pi$ &11.07  &$\rho\eta(1475)$ & 0.17 \\
                              &$\omega(1650)\pi$  &0.11  &$\omega\pi(1300)$ & 1.02   \\
                              &$KK^*$  &0.02  &$\rho\eta(1295)$ & 0.67 \\
                              &$KK^*(1410)$  &$\textless0.01$   &$\rho(1450)\eta$ & 0.65   \\
                              &$\rho\eta'$ &0.14  & &     \\
                               \hline
  $1^-\rightarrow 1^-1^-$     &$\rho\rho$   &5.64   &$K^*K^*$  &0.15\\
                                &$\rho\rho(1450)$   &12.93   &$K^*K^*(1410)$  &0.49   \\
                               \hline
  $1^-\rightarrow 0^-1^+$    &$a_{1}(1260)\pi$   &3.26   &$b_{1}(1235)\eta'$  &0.17    \\
                              &$h_{1}(1170)\pi$   &4.22   &$\pi a_1(1640)$  &9.81 \\
                              &$KK_{1}(1400)$  &0.10   &$b_{1}(1235)\eta$  &0.49 \\
                               &$KK_{1}(1270)$  &0.10&  &  \\
                                 \hline
  $1^-\rightarrow 0^-2^+$     &$a_{2}(1320)\pi$  &4.83 &$a_{2}(1700)\pi$  &27.80 \\
                              &$KK_{2}^*(1430)$ &0.01&  & \\
                               \hline
  $1^-\rightarrow 0^- 2^-$            &$\pi\pi_{2}(1670)$ &3.65 &$\pi\eta_{2}(1645)$ &1.94\\
                                      &$KK_2(1770)$ &0.02 &$KK_2(1820)$ &0.03\\
   \hline
  $1^-\rightarrow 0^- 3^-$            &$\pi\omega_{3}(1670)$ &2.46 &$\eta\rho_{3}(1690)$ &0.02 \\
                                      &$KK_{3}(1780)$ &$\textless0.01$&   &  \\
   \hline
  $1^-\rightarrow 1^- 1^+$            &$b_{1}(1235)\rho$  &3.38 &$K_1(1400)K^*$&0.02\\
                                      &$a_{1}(1260)\omega$  &3.14  &$K_1(1270)K^*$&0.09\\
                                      &$\rho f_1(1285)$&3.00 &$\rho f_1(1420)$&0.16\\
                                      &$\rho h_1(1170)$&2.50&  & \\
                                       \hline
     $1^-\rightarrow 2^+ 1^-$          &$\rho f_2(1270)$&3.93&$K^*K_2^*(1430)$&0.03\\
                                         &$\omega a_2(1320)$&5.10&  & \\
                                         \hline
         $1^-\rightarrow 0^- 4^+$        & $a_4(1970)\pi$&0.91&  &   \\
         \hline
           $1^-\rightarrow 1^- 0^+$          &$a_0(1450)\omega$&0.52&  &   \\
           \hline

Total width    &\multicolumn{4}{c}{143.01}\\
\hline\hline
\end{tabular}
\end{center}
\end{table}

\begin{table}[htpb]
\begin{center}
\caption{ \label{tab:4D}The decay widths of $\rho(4^3D_1)$ (in MeV),  the initial state  mass is set to be 2624 MeV and the masses of all the final states are taken from PRR~\cite{Zyla:2020zbs}.}
\begin{tabular}{c|cc|cc}
\hline\hline
 Channel                 & Mode      &$\rho(4^3D_1)$            & Mode           &$\rho(4^3D_1)$   \\
\hline
 $1^-\rightarrow 0^-0^-$      & $\pi\pi$     & 5.55  & $KK$  &$\textless0.01$ \\
                              & $\pi\pi(1300)$  & 11.16 & $KK(1460)$  & $\textless0.01$  \\
                               & $\pi\pi(1800)$  &7.07  &  &    \\
  \hline
  $1^-\rightarrow 0^-1^-$     &$\pi\omega$  &0.48       &$KK^*(1680)$ & 0.01 \\
                              &$\rho\eta$  &0.07    &$\rho(1700)\eta$ & 0.07\\
                              &$\omega(1420)\pi$ &2.23  &$\rho\eta(1475)$ &$\textless0.01$  \\
                              &$\omega(1650)\pi$  &0.23  &$\omega\pi(1300)$ & 0.61   \\
                              &$KK^*$  &0.01  &$\rho\eta(1295)$ & 0.36 \\
                              &$KK^*(1410)$  &$\textless0.01$   &$\rho(1450)\eta$ & 0.37  \\
                              &$\rho\eta'$ &0.14 & &     \\
                               \hline
  $1^-\rightarrow 1^-1^-$     &$\rho\rho$   &4.81  &$K^*K^*$  &0.02\\
                                &$\rho\rho(1450)$   &14.42  &$K^*K^*(1410)$  &0.06  \\
                               \hline
  $1^-\rightarrow 0^-1^+$    &$a_{1}(1260)\pi$   &4.00   &$b_{1}(1235)\eta'$  &0.10   \\
                              &$h_{1}(1170)\pi$   &6.90   &$\pi a_1(1640)$  &10.54\\
                              &$KK_{1}(1400)$  &0.04  &$b_{1}(1235)\eta$  &0.82 \\
                               &$KK_{1}(1270)$  &0.05&  &  \\
                                 \hline
  $1^-\rightarrow 0^-2^+$     &$a_{2}(1320)\pi$  &0.96 &$a_{2}(1700)\pi$  &7.09 \\
                              &$KK_{2}^*(1430)$ &$\textless0.01$&  & \\
                               \hline
  $1^-\rightarrow 0^- 2^-$            &$\pi\pi_{2}(1670)$ &6.02 &$\pi\eta_{2}(1645)$ &4.48\\
                                      &$KK_2(1770)$ &0.05 &$KK_2(1820)$ &0.01\\
   \hline
  $1^-\rightarrow 0^- 3^-$            &$\pi\omega_{3}(1670)$ &1.35 &$\eta\rho_{3}(1690)$ &0.10 \\
                                      &$KK_{3}(1780)$ &$\textless0.01$&   &  \\
   \hline
  $1^-\rightarrow 1^- 1^+$            &$b_{1}(1235)\rho$  &1.15 &$K_1(1400)K^*$&0.04\\
                                      &$a_{1}(1260)\omega$  &0.39  &$K_1(1270)K^*$&0.08\\
                                      &$\rho f_1(1285)$&0.52 &$\rho f_1(1420)$&0.14\\
                                      &$\rho h_1(1170)$&1.14&  & \\
                                       \hline
     $1^-\rightarrow 2^+ 1^-$          &$\rho f_2(1270)$&5.71&$K^*K_2^*(1430)$&0.01\\
                                         &$\omega a_2(1320)$&3.58&  & \\
                                         \hline
         $1^-\rightarrow 0^- 4^+$        & $a_4(1970)\pi$&0.40&  &   \\
         \hline
           $1^-\rightarrow 1^- 0^+$          &$a_0(1450)\omega$&0.60&  &   \\
           \hline

Total width    &\multicolumn{4}{c}{105.82}\\
\hline\hline
\end{tabular}
\end{center}
\end{table}

\begin{table}[htpb]
\begin{center}
\caption{ \label{tab:6S}The decay widths of $\rho(6^3S_1)$ (in MeV),  the initial state mass is set to be 2774 MeV and the masses of all the final states are taken from RPP~\cite{Zyla:2020zbs}.}
\begin{tabular}{c|cc|cc}
\hline\hline
 Channel                 & Mode      &$\rho(6^3S_1)$            & Mode           &$\rho(6^3S_1)$   \\
\hline
 $1^-\rightarrow 0^-0^-$      & $\pi\pi$     & 0.51  & $KK$  &$\textless0.01$ \\
                              & $\pi\pi(1300)$  & 4.83 & $KK(1460)$  &$\textless0.01$   \\
                               & $\pi\pi(1800)$  &5.36  &  &    \\
  \hline
  $1^-\rightarrow 0^-1^-$     &$\pi\omega$  &$\textless0.01$        &$KK^*(1680)$ &$\textless0.01$ \\
                              &$\rho\eta$  &0.02   &$\rho(1700)\eta$ & 0.04\\
                              &$\omega(1420)\pi$ &3.29  &$\rho\eta(1475)$ &0.12  \\
                              &$\omega(1650)\pi$  &0.03  &$\omega\pi(1300)$ &0.21   \\
                              &$KK^*$  &$\textless0.01$  &$\rho\eta(1295)$ &0.14\\
                              &$KK^*(1410)$  &0.01   &$\rho(1450)\eta$ &0.10  \\
                              &$\rho\eta'$ &0.08 & &     \\
                               \hline
  $1^-\rightarrow 1^-1^-$     &$\rho\rho$   &3.67 &$K^*K^*$  &$\textless0.01$\\
                                &$\rho\rho(1450)$   &7.26  &$K^*K^*(1410)$  &0.10 \\
                               \hline
  $1^-\rightarrow 0^-1^+$    &$a_{1}(1260)\pi$   &0.59  &$b_{1}(1235)\eta'$  &0.09   \\
                              &$h_{1}(1170)\pi$   &0.79  &$\pi a_1(1640)$  &5.44\\
                              &$KK_{1}(1400)$  &0.02  &$b_{1}(1235)\eta$  &0.08 \\
                               &$KK_{1}(1270)$  &0.02&  &  \\
                                 \hline
  $1^-\rightarrow 0^-2^+$     &$a_{2}(1320)\pi$  &0.36 &$a_{2}(1700)\pi$  &13.82 \\
                              &$KK_{2}^*(1430)$ &$\textless0.01$&  & \\
                               \hline
  $1^-\rightarrow 0^- 2^-$            &$\pi\pi_{2}(1670)$ &0.95 &$\pi\eta_{2}(1645)$ &0.59\\
                                      &$KK_2(1770)$ &$\textless0.01$ &$KK_2(1820)$ &$\textless0.01$\\
   \hline
  $1^-\rightarrow 0^- 3^-$            &$\pi\omega_{3}(1670)$ &0.63 &$\eta\rho_{3}(1690)$ &$\textless0.01$ \\
                                      &$KK_{3}(1780)$ &$\textless0.01$&   &  \\
   \hline
  $1^-\rightarrow 1^- 1^+$            &$b_{1}(1235)\rho$  &2.57 &$K_1(1400)K^*$&$\textless0.01$\\
                                      &$a_{1}(1260)\omega$  &1.78  &$K_1(1270)K^*$&0.01\\
                                      &$\rho f_1(1285)$&1.68 &$\rho f_1(1420)$&0.10\\
                                      &$\rho h_1(1170)$&1.94&  & \\
                                       \hline
     $1^-\rightarrow 2^+ 1^-$          &$\rho f_2(1270)$&4.12&$K^*K_2^*(1430)$&$\textless0.01$\\
                                         &$\omega a_2(1320)$&4.36&  & \\
                                         \hline
         $1^-\rightarrow 0^- 4^+$        & $a_4(1970)\pi$&1.31&  &   \\
         \hline
           $1^-\rightarrow 1^- 0^+$          &$a_0(1450)\omega$&0.36&  &   \\
           \hline

Total width    &\multicolumn{4}{c}{67.46}\\
\hline\hline
\end{tabular}
\end{center}
\end{table}

\begin{table}[htpb]
\begin{center}
\caption{ \label{tab:5D}The decay widths of $\rho(5^3D_1)$ (in MeV),  the initial state  mass is set to be 2840 MeV and the masses of the final states are taken from RPP~\cite{Zyla:2020zbs}.}
\begin{tabular}{c|cc|cc}
\hline\hline
 Channel                 & Mode      &$\rho(5^3D_1)$            & Mode           &$\rho(5^3D_1)$   \\
\hline
 $1^-\rightarrow 0^-0^-$      & $\pi\pi$     & 3.46  & $KK$  &$\textless0.01$ \\
                              & $\pi\pi(1300)$  & 10.25 & $KK(1460)$  & $\textless0.01$  \\
                               & $\pi\pi(1800)$  &7.90  &  &    \\
  \hline
  $1^-\rightarrow 0^-1^-$     &$\pi\omega$  &0.04       &$KK^*(1680)$ &0.10\\
                              &$\rho\eta$  &$\textless0.01$  &$\rho(1700)\eta$ & 0.07\\
                              &$\omega(1420)\pi$ &1.19  &$\rho\eta(1475)$ &0.01 \\
                              &$\omega(1650)\pi$  &0.20 &$\omega\pi(1300)$ &0.17   \\
                              &$KK^*$  &$\textless0.01$  &$\rho\eta(1295)$ &0.11\\
                              &$KK^*(1410)$  &0.02   &$\rho(1450)\eta$ &0.10  \\
                              &$\rho\eta'$ &0.04 & &     \\
                               \hline
  $1^-\rightarrow 1^-1^-$     &$\rho\rho$   &0.58 &$K^*K^*$  &$\textless0.01$\\
                                &$\rho\rho(1450)$   &5.44 &$K^*K^*(1410)$  &0.04 \\
                               \hline
  $1^-\rightarrow 0^-1^+$    &$a_{1}(1260)\pi$   &1.34  &$b_{1}(1235)\eta'$  &0.07   \\
                              &$h_{1}(1170)\pi$   &2.64  &$\pi a_1(1640)$  &7.35\\
                              &$KK_{1}(1400)$  &0.02  &$b_{1}(1235)\eta$  &0.16 \\
                               &$KK_{1}(1270)$  &0.02&  &  \\
                                 \hline
  $1^-\rightarrow 0^-2^+$     &$a_{2}(1320)\pi$  &0.84 &$a_{2}(1700)\pi$  &5.81 \\
                              &$KK_{2}^*(1430)$ &$\textless0.01$&  & \\
                               \hline
  $1^-\rightarrow 0^- 2^-$            &$\pi\pi_{2}(1670)$ &1.58&$\pi\eta_{2}(1645)$ &1.68\\
                                      &$KK_2(1770)$ &0.01 &$KK_2(1820)$ &$\textless0.01$\\
   \hline
  $1^-\rightarrow 0^- 3^-$            &$\pi\omega_{3}(1670)$ &0.62 &$\eta\rho_{3}(1690)$ &0.03 \\
                                      &$KK_{3}(1780)$ &$\textless0.01$&   &  \\
   \hline
  $1^-\rightarrow 1^- 1^+$            &$b_{1}(1235)\rho$  &1.22 &$K_1(1400)K^*$&0.01\\
                                      &$a_{1}(1260)\omega$  &0.69 &$K_1(1270)K^*$&0.03\\
                                      &$\rho f_1(1285)$&0.91 &$\rho f_1(1420)$&0.12\\
                                      &$\rho h_1(1170)$&0.93&  & \\
                                       \hline
     $1^-\rightarrow 2^+ 1^-$          &$\rho f_2(1270)$&1.41&$K^*K_2^*(1430)$&$\textless0.01$\\
                                         &$\omega a_2(1320)$&0.84&  & \\
                                         \hline
         $1^-\rightarrow 0^- 4^+$        & $a_4(1970)\pi$&0.52&  &   \\
         \hline
           $1^-\rightarrow 1^- 0^+$          &$a_0(1450)\omega$&0.46&  &   \\
           \hline

Total width    &\multicolumn{4}{c}{59.00}\\
\hline\hline
\end{tabular}
\end{center}
\end{table}

\begin{table}[htpb]
\begin{center}
\caption{ \label{tab:7S}The decay widths of $\rho(7^3S_1)$ (in MeV),  the initial  state mass is set to be 2967 MeV and the masses of  the final states are taken from RPP~\cite{Zyla:2020zbs}.}
\begin{tabular}{c|cc|cc}
\hline\hline
 Channel                 & Mode      &$\rho(7^3S_1)$            & Mode           &$\rho(7^3S_1)$   \\
\hline
 $1^-\rightarrow 0^-0^-$      & $\pi\pi$     & $\textless0.01$ & $KK$  &$\textless0.01$ \\
                              & $\pi\pi(1300)$  & 4.24 & $KK(1460)$  & 0.01  \\
                               & $\pi\pi(1800)$  &5.77  &  &    \\
  \hline
  $1^-\rightarrow 0^-1^-$     &$\pi\omega$  & 0.16      &$KK^*(1680)$ &$\textless0.01$\\
                              &$\rho\eta$  &0.16 &$\rho(1700)\eta$ &0.09 \\
                              &$\omega(1420)\pi$ &1.14  &$\rho\eta(1475)$ &0.12 \\
                              &$\omega(1650)\pi$  &0.16 &$\omega\pi(1300)$ & $\textless0.01$ \\
                              &$KK^*$  &$\textless0.01$ &$\rho\eta(1295)$ &$\textless0.01$   \\
                              &$KK^*(1410)$  &$\textless0.01$  &$\rho(1450)\eta$ & $\textless0.01$\\
                              &$\rho\eta'$ &0.11 & &     \\
                               \hline
  $1^-\rightarrow 1^-1^-$     &$\rho\rho$   &6.86    &$K^*K^*$  &0.04\\
                                &$\rho\rho(1450)$   &9.09    &$K^*K^*(1410)$  &$\textless0.01$ \\
                               \hline
  $1^-\rightarrow 0^-1^+$    &$a_{1}(1260)\pi$   & 1.21 &$b_{1}(1235)\eta'$  & 0.08 \\
                              &$h_{1}(1170)\pi$   &0.42 &$\pi a_1(1640)$  &4.28\\
                              &$KK_{1}(1400)$  & 0.01&$b_{1}(1235)\eta$  &0.15\\
                               &$KK_{1}(1270)$  &0.01&  &  \\
                                 \hline
  $1^-\rightarrow 0^-2^+$     &$a_{2}(1320)\pi$  &$\textless0.01$ &$a_{2}(1700)\pi$  &6.86 \\
                              &$KK_{2}^*(1430)$ &0.02    &  & \\
                               \hline
  $1^-\rightarrow 0^- 2^-$            &$\pi\pi_{2}(1670)$ &0.70   &$\pi\eta_{2}(1645)$ &0.55\\
                                      &$KK_2(1770)$ &0.01&$KK_2(1820)$ &\\
   \hline
  $1^-\rightarrow 0^- 3^-$            &$\pi\omega_{3}(1670)$ &0.04 &$\eta\rho_{3}(1690)$ &0.05    \\
                                      &$KK_{3}(1780)$ &\textless0.01  &   &  \\
   \hline
  $1^-\rightarrow 1^- 1^+$            &$b_{1}(1235)\rho$  &3.09 &$K_1(1400)K^*$&$\textless0.01$\\
                                      &$a_{1}(1260)\omega$  &1.92   &$K_1(1270)K^*$&$\textless0.01$    \\
                                      &$\rho f_1(1285)$&1.57 &$\rho f_1(1420)$&0.06     \\
                                      &$\rho h_1(1170)$&3.21     &  & \\
                                       \hline
     $1^-\rightarrow 2^+ 1^-$          &$\rho f_2(1270)$& 4.21   &$K^*K_2^*(1430)$&0.02  \\
                                         &$\omega a_2(1320)$&3.72 &  & \\
                                         \hline
         $1^-\rightarrow 0^- 4^+$        & $a_4(1970)\pi$&0.53 &  &   \\
         \hline
           $1^-\rightarrow 1^- 0^+$          &$a_0(1450)\omega$& 0.24 &  &   \\
           \hline

Total width    &\multicolumn{4}{c}{61.37 }\\
\hline\hline
\end{tabular}
\end{center}
\end{table}

\begin{table}[htpb]
\begin{center}
\caption{ \label{tab:6D}The decay widths of $\rho(6^3D_1)$ (in MeV),  the initial  state mass is set to be 3020 MeV and the masses of  the final states are taken from RPP~\cite{Zyla:2020zbs}.}
\begin{tabular}{c|cc|cc}
\hline\hline
 Channel                 & Mode      &$\rho(6^3D_1)$            & Mode           &$\rho(6^3D_1)$   \\
\hline
 $1^-\rightarrow 0^-0^-$      & $\pi\pi$     & 0.84 & $KK$  &$\textless0.01$ \\
                              & $\pi\pi(1300)$  &4.29 & $KK(1460)$  & $\textless0.01$  \\
                               & $\pi\pi(1800)$  &5.22  &  &    \\
  \hline
  $1^-\rightarrow 0^-1^-$     &$\pi\omega$  & $\textless0.01$     &$KK^*(1680)$ &$\textless0.01$\\
                              &$\rho\eta$  &0.02 &$\rho(1700)\eta$ &0.03 \\
                              &$\omega(1420)\pi$ &0.40  &$\rho\eta(1475)$ &0.03 \\
                              &$\omega(1650)\pi$  &0.06 &$\omega\pi(1300)$ & 0.02 \\
                              &$KK^*$  &$\textless0.01$ &$\rho\eta(1295)$ &0.01   \\
                              &$KK^*(1410)$  &0.01 &$\rho(1450)\eta$ & $\textless0.01$\\
                              &$\rho\eta'$ &0.03 & &     \\
                               \hline
  $1^-\rightarrow 1^-1^-$     &$\rho\rho$   &0.37    &$K^*K^*$  &$\textless0.01$\\
                                &$\rho\rho(1450)$   &0.47    &$K^*K^*(1410)$  &0.03 \\
                               \hline
  $1^-\rightarrow 0^-1^+$    &$a_{1}(1260)\pi$   & 0.16 &$b_{1}(1235)\eta'$  & 0.05 \\
                              &$h_{1}(1170)\pi$   &0.40 &$\pi a_1(1640)$  &2.90\\
                              &$KK_{1}(1400)$  & $\textless0.01$&$b_{1}(1235)\eta$  &$\textless0.01$\\
                               &$KK_{1}(1270)$  &$\textless0.01$&  &  \\
                                 \hline
  $1^-\rightarrow 0^-2^+$     &$a_{2}(1320)\pi$  &0.05 &$a_{2}(1700)\pi$  &2.93 \\
                              &$KK_{2}^*(1430)$ &$\textless0.01$    &  & \\
                               \hline
  $1^-\rightarrow 0^- 2^-$            &$\pi\pi_{2}(1670)$ &0.14   &$\pi\eta_{2}(1645)$ &0.16\\
                                      &$KK_2(1770)$ &$\textless0.01$&$KK_2(1820)$ &$\textless0.01$\\
   \hline
  $1^-\rightarrow 0^- 3^-$            &$\pi\omega_{3}(1670)$ &0.10 &$\eta\rho_{3}(1690)$ &$\textless0.01$    \\
                                      &$KK_{3}(1780)$ &$\textless0.01$&   &  \\
   \hline
  $1^-\rightarrow 1^- 1^+$            &$b_{1}(1235)\rho$  &0.91 &$K_1(1400)K^*$&$\textless0.01$\\
                                      &$a_{1}(1260)\omega$  &0.97  &$K_1(1270)K^*$&$\textless0.01$    \\
                                      &$\rho f_1(1285)$&0.98 &$\rho f_1(1420)$&0.07     \\
                                      &$\rho h_1(1170)$&0.77    &  & \\
                                       \hline
     $1^-\rightarrow 2^+ 1^-$          &$\rho f_2(1270)$&0.37   &$K^*K_2^*(1430)$&$\textless0.01$  \\
                                         &$\omega a_2(1320)$&0.05 &  & \\
                                         \hline
         $1^-\rightarrow 0^- 4^+$        & $a_4(1970)\pi$&0.31 &  &   \\
         \hline
           $1^-\rightarrow 1^- 0^+$          &$a_0(1450)\omega$& 0.27 &  &   \\
           \hline

Total width    &\multicolumn{4}{c}{23.88 }\\
\hline\hline
\end{tabular}
\end{center}
\end{table}

In this work, we employ the $^3P_0$ model with the realistic meson wave functions obtained from the MGI model to evaluate the decay widths of $\rho(5^3S_1)$, $\rho(4^3D_1)$, $\rho(6^3S_1)$, $\rho(5^3D_1)$, $\rho(7^3S_1)$, and $\rho(6^3D_1)$. The value of parameter $\gamma$ is taken from our previous work~\cite{Li:2021qgz}, 
where it was obtained by fitting to the decay widths of  the light mesons with same quantum numbers as $\rho$ mesons.
 $\gamma=6.57$ is used for the $u\bar{u}$/$d\bar{d}$ pair creation, 
 and the $\gamma$ value  of $s\bar{s}$ pair creation  is suppressed by an additional factor $m_u/m_s$.
The predicted masses by the MGI model are used 
for the unobserved excited $\rho$ mesons in the calculations of the strong decay widths, and the masses of  the other mesons are taken from Review of Particle Physics (RPP)~\cite{Zyla:2020zbs}.

The partial widths and total width of $\rho(5^3S_1)$ are listed in Table~\ref{tab:5S}. The total decay width of $\rho(5^3S_1)$ is expected to be $\Gamma=143$~MeV, which is close to the lower limit of the width $\Gamma=(209\pm26)$~MeV obtained in Ref.~\cite{Lichard:2018enc}. The main
decay modes of $\rho(5^3S_1)$ are $\pi\pi(1300)(\to 4\pi)$, $\pi\pi(1800)(\to 4\pi)$, $\pi\omega(1420)(\to 4\pi, 6\pi)$, $\rho\rho(1450)(\to 4\pi, 6\pi)$, $\pi a_1(1640)(\to 4\pi)$, and $\pi a_2(1700)(\to \eta\pi\pi, \pi K\bar{K})$, thus one can search for the $\rho(5^3S_1)$ in the channels of $4\pi$, $6\pi$, $\eta\pi\pi$, and $\pi K\bar{K}$.

 The partial widths and total width of $\rho(4^3D_1)$ are listed in Table~\ref{tab:4D}. The total decay width of $\rho(4^3D_1)$ is expected to be $\Gamma=105.82$~MeV, consistent with the broad structure around 2.64~GeV in Fig.~\ref{fig:zhiliang}. The predicted main
  decay modes are $\pi\pi(1300)(\to 4\pi)$,  $\pi\pi(1800)(\to 4\pi)$, $\rho\rho(1450)(\to 4\pi)$, $\pi h_1(1170)(\to 4\pi)$, $\pi a_1(1640)(\to 4\pi)$, and $\pi\pi_1(1670)(\to 4\pi)$, thus one can search for $\rho(4^3D_1)$ in the $4\pi$ channel.  

 The partial widths and total width of $\rho(6^3S_1)$ are listed in Table~\ref{tab:6S}. The total decay width of $\rho(6^3S_1)$ is expected to be $\Gamma=67.46$~MeV, and the main decay modes are predicted to be $\pi\pi(1300)(\to 4\pi)$,  $\pi\pi(1800)(\to 4\pi)$, $\rho\rho(1450)(\to 4\pi, 6\pi)$, $\pi a_2(1700)(\to \eta\pi\pi, \pi K\bar{K})$, $\rho f_2(1270)(\to 4\pi)$, and $\omega a_2(1320)$. Considering the width of the intermediate state $a_2(1320)$ is $107\pm 5$~MeV, the $\rho(6^3S_1)$ state is expected to be observed in the $\omega a_2(1320)$ channel.
 
The partial widths and total width of $\rho(5^3D_1)$ are listed in Table~\ref{tab:5D}. The total decay width of $\rho(5^3D_1)$ is expected to be $\Gamma=59.00$~MeV, and the predicted main decay modes are $\pi\pi(1300)(\to 4\pi)$, $\pi\pi(1800)(\to 4\pi)$, $\rho\rho(1450)(\to 4\pi)$,  and $\pi a_1(1640)(\to 4\pi)$. Thus $\rho(5^3D_1)$ is expected to be observed in $4\pi$ channel.

The partial widths and total width of $\rho(7^3S_1)$ are listed in Table~\ref{tab:7S}. The total decay width of $\rho(7^3S_1)$ is expected to be $\Gamma=61.37$~MeV, and the predicted main decay modes are $\pi\pi(1300)(\to 4\pi)$,  $\pi\pi(1800)(\to 4\pi)$, $\rho\rho(\to 4\pi)$, $\rho\rho(1450)(\to 4\pi, 6\pi)$,  $\pi a_2(1700)(\to 4\pi)$, $\pi a_1(1640)(\to 4\pi)$, and $\omega a_2(1320)$. Thus one can search for the $\rho(7^3S_1)$ state in the $\omega a_2(1320)$ and $4\pi$ channels.

The partial widths and total width of $\rho(6^3D_1)$ are listed in Table~\ref{tab:6D}. The total decay width of $\rho(6^3D_1)$ is expected to be $\Gamma=23.88$~MeV, and the predicted main decay modes are $\pi\pi(1300)(\to 4\pi)$,  $\pi\pi(1800)(\to 4\pi)$, $\pi a_1(1640)(\to 4\pi)$, and $\pi a_2(1700)(\to 4\pi)$. The state $\rho(6^3D_1)$ is expected to be observed in the $4\pi$ channel.
 
\section{SUMMARY AND CONCLUSION}
\label{sec5}
In this work, we  present the masses and strong decay widths of the higher excited $\rho$ 
mesons
$\rho(5^3S_1)$, $\rho(4^3D_1)$, $\rho(6^3S_1)$, $\rho(5^3D_1)$, $\rho(7^3S_1)$, and $\rho(6^3D_1)$. The deviations between the theoretical masses predicted with MGI and GI models are about $300\sim 500$~MeV, which implies the screening effects
are important for the higher excited $\rho$ mesons. We   also show
some experimental hints for the existences of the higher excited $\rho$ mesons, and one can not claim these resonances based on the poor measurements at present.

In addition to the masses of the higher exited $\rho$ mesons, the strong decay widths are also calculated, which are $\Gamma_{\rho(5S)}=143$~MeV,  $\Gamma_{\rho(4D)}=106$~MeV, $\Gamma_{\rho(6S)}=67$~MeV, $\Gamma_{\rho(5D)}=59$~MeV,  $\Gamma_{\rho(7S)}=61$~MeV, and $\Gamma_{\rho(6D)}=24$~MeV, respectively. According 
to our theoretical predictions, those states are expected to be observed in the final states of $4\pi$, $6\pi$, $\eta \pi\pi$, $\pi K\bar{K}$, $\omega a_2(1320)$.  
Our theoretical study on
their masses and decay properties should be useful to search for these resonances experimentally.
\section*{Acknowledgements}

This work is supported by the Natural Science Foundation of Henan under Grand No. 222300420554. It is also supported by the Key Research Projects of Henan Higher Education Institutions under No. 20A140027, the Project of Youth Backbone Teachers of Colleges and Universities of Henan Province (2020GGJS017), the Youth Talent Support Project of Henan (2021HYTP002), the Fundamental Research Cultivation Fund for Young Teachers of Zhengzhou University (JC202041042), and the Open Project of Guangxi Key Laboratory of Nuclear Physics and Nuclear Technology, No.NLK2021-08.

\end{document}